\documentclass[pre,amssymb,twocolumn,raggedbottom]{revtex4}
     \usepackage{graphicx}

\begin{document}

\title{Different thermodynamic pathways to the solvation free energy of a
spherical cavity in a hard sphere fluid}
\author{Yng-gwei Chen$^{1,3}$}
\author{John D. Weeks$^{2,3}$}
\affiliation{
$^{1}$Department of Physics, $^{2}$Department of Chemistry and Biochemistry,
and $^{3}$Institute for Physical Science and Technology,
University of Maryland, College Park, Maryland 20742}

\date{\today}

\begin{abstract}
This paper determines the excess free energy associated with the formation
of a spherical cavity in a hard sphere fluid. The solvation free energy can
be calculated by integration of the structural changes induced by inserting
the cavity using a number of different exact thermodynamic pathways. We
consider three such pathways, including a new density route derived here.
Structural information about the nonuniform hard sphere fluid in the
presence of a general external field is given by the recently developed
hydrostatic linear response (HLR) integral equation. Use of the HLR results
in the different pathways gives a generally accurate determination of the
solvation free energy for cavities over a wide range of sizes, from zero to
infinity. Results for a related method, the Gaussian Field Model, are also
discussed.
\end{abstract}

\maketitle

\section{Introduction}

The solvation free energy determines how readily a solute can be dissolved
in a given solvent fluid. This plays an important role in many chemically
and biologically important processes, perhaps most notably in hydrophobic
interactions in water. A significant part of the solvation free energy
arises from the required expulsion of solvent molecules from the region
occupied by the harshly repulsive molecular core of the solute. These very
strong ``excluded volume'' interactions can significantly perturb the local
density around the solute and cause simple approaches based on gradient
expansions to fail.

These effects can be seen most clearly in the simple model system treated in
this paper. We will calculate the excess or solvation free energy associated
with the insertion of a spherical cavity with radius $R_{v}$ into a hard
sphere fluid, whose molecules have diameter $\sigma $. By definition, the
centers of the solvent molecules are completely excluded from the region of
the cavity, which thus acts like a hard core external field.

This system has many interesting limits. When the exclusion field or cavity
radius $R_{v}$ equals $\sigma $, then the cavity acts like another solvent
particle and the solvation free energy is directly related to the chemical
potential of the solvent. As the cavity radius tends to infinity it
effectively turns into a hard wall and the relevant thermodynamic quantity
is the surface free energy or surface tension associated with a hard wall in
a hard sphere fluid. A cavity with radius $R_{v}=\sigma /2$ acts like a hard
core ``point solute'' of zero diameter. Even shorter-ranged hard core fields
or {\em ``tiny cavities''} with $R_{v}\leq \sigma /2$ are also of interest,
since the induced structure and solvation free energy of a tiny cavity can
be calculated exactly. This limit can thus serve as a nontrivial check on
approximate methods.

The most commonly used method today for such problems is weighted density
functional theory (DFT) \cite{evans:density}. Here one attempts to
describe the free energy
directly as a functional of some kind of smoothed or weighted average of the
nonuniform and often rapidly varying singlet density. This has the advantage
that the free energy is obtained directly and by construction the associated
fluid structure (obtained by functionally differentiating the free energy)
is consistent with the approximate free energy. However the choice of
appropriate weighting functions is by no means obvious and a number of
different and often highly formal schemes have been proposed.

We focus instead in this paper on making direct use of structural
information about the nonuniform solvent fluid to obtain the solvation free
energy. We believe this allows physical intuition to play a more central
role and we can take advantage of the recent development of a generally very
accurate theory relating the structure of a nonuniform hard sphere fluid to
the associated external field \cite{hlr}.

As we will see below, the free energy can then be calculated by integration,
starting from an initially known state (e.g., the uniform fluid) and
determining the free energy changes as the solute-solvent interaction (the
hard core external field) is ``turned on'', or alternatively, as the density
is changed from the initial to the final state. There exist many possible
routes from the initial to the final state, and we will generally refer to
them as {\em thermodynamic pathways}. If exact results are used for the
intermediate values of the structure and associated fields, then all these
different pathways will give the same (exact) result for the free energy.

In practice, of course, approximations will have to be made and the
different pathways will generally yield different results. This is
sometimes referred to as the ``thermodynamic inconsistency'' of
structurally based methods \cite{evans:density}. But this can be viewed
more positively as giving one the freedom to choose particular pathways
that could be relatively insensitive to the errors that exist in the
structural theory, and we will try to
use this flexibility to obtain the most accurate
results. Moreover, there is an inherent smoothing of the structural
information in the integration used to obtain the free energy. The
differences in free energy predicted by different pathways will also give
us some indication about the overall quality of the theory.

This approach generally requires the density profiles and associated fields of
all the intermediate states along the various pathways, and thus a fast and
accurate method for determining these quantities is crucial for computation
efficiency. We will use here the generally accurate {\em hydrostatic linear
response} (HLR) equation \cite{hlr} proposed by Katsov and Weeks. A different
physically motivated derivation of the HLR equation is given below.

We will also examine the alternative free energy predictions that arise from
a theory closely related to the HLR equation, the {\em Gaussian field
model }(GFM) developed by Chandler \cite{gft}. For a solute with a hard core
the GFM proposes an approximate partition function from which the associated
density response can be derived. In the particular case where a rigid cavity
is inserted into a hard sphere fluid, the HLR and the GFM approaches turn
out to make identical predictions for the induced structure. Thus
structurally based routes to the free energy involving only hard core fields
will give the same results. In addition, one can use the approximate
GFM partition function to evaluate the solvation free energy 
directly. However, as we will
show later, the latter approach tends to produce less accurate results. This
deficiency shows up even more strongly in the tiny cavity limit where the
structural predictions of the HLR and the GFM are {\em exact}, and several
pathways giving the exact free energy can be found. This illustrates the
advantage of considering a variety of thermodynamic pathways that can make
best use of the available structural information.

\section{Density response to an external field}

\subsection{ The HLR equation}

\label{sec hlr}

We describe the system using a grand canonical ensemble, and thus want to
determine the excess grand free energy arising from insertion of a spherical
cavity or hard core external field of radius $R_{v}$. To derive the HLR
equation \cite{hlr} we start with the basic linear
response equation \cite{hansenmac} for a nonuniform
hard sphere system in a {\em general} external field $\phi ({\bf r})$, with
chemical potential $\mu ^{B}$, inverse temperature $\beta =(k_{B}T)^{-1}$
and associated density $\rho ({\bf r};\ \mu ^{B},[\phi ])\equiv $ $\rho (%
{\bf r}{\bf )}$:
\begin{equation}
-\beta \delta \phi ({\bf r}_{1})=\int \!d{\bf r}_{2}\,\chi ^{-1}({\bf r}_{1},%
{\bf r}_{2};{\bf [}\rho ])\delta \rho ({\bf r}_{2}).  \label{linresponse}
\end{equation}
This relates {\em small} perturbations in the density and field through the
(inverse) linear response function
\begin{equation}
\chi ^{-1}({\bf r}_{1},{\bf r}_{2};{\bf [}\rho ])\equiv \delta ({\bf r}%
_{1}\!-\!{\bf r}_{2})/\rho ({\bf r}_{1})\!-\!c({\bf r}_{1},{\bf r}_{2};{\bf [%
}\rho ]).  \label{chiinverse}
\end{equation}
Here $c({\bf r}_{1},{\bf r}_{2};{\bf [}\rho ])$ is the direct correlation
function of the nonuniform hard sphere system$.$ The notation $[\rho ]$
indicates that these correlation functions are nonlocal functionals of the
density $\rho ({\bf r}).$

Since we want to focus on the effects of the perturbing field, we have used
the {\em inverse} form of linear response theory \cite{prattotf}
in Eq.\ (\ref{linresponse}), where the field appears explicitly only on the left
hand side, evaluated at ${\bf r}_{1}.$ This provides many advantages
in dealing with large field perturbations,
as will soon become apparent. In most cases we will consider
perturbations about a {\em uniform} system with chemical potential $\mu $ and
density $\rho (\mu ) \equiv \rho ({\bf r};\ \mu ,[0])$.
When using this simplified notation $\rho (\mu )$ should not be confused
with $\rho ({\bf r}{\bf )}\equiv \rho ({\bf r};\ \mu ^{B},[\phi ])$.
Similarly, we will let $\mu (\rho )$ denote the chemical
potential of the uniform fluid as a function of density $\rho $. In a
uniform system the direct correlation function $\!c$ will take the simple
form $\!c(r_{12};\rho ),$ where $r_{12}\equiv |{\bf r}_{1}-{\bf r}_{2}|$.

But how can we use Eq.\ (\ref{linresponse}) to describe the density response to
a {\em large} field perturbation such as the hard core field of interest
here? This linear relation between a (possibly infinite) external field
perturbation on the left hand side and the finite induced density change on
the right must certainly fail for values of ${\bf r}_{1}$ where the field is
very large. Conversely, Eq.\ (\ref{linresponse}) should be most accurate for
those values of ${\bf r}_{1}$ where the field is small --- in particular
where the field {\em vanishes} --- and then through the integration over all
${\bf r}_{2}$ it relates density changes in regions where the field vanishes
to density changes in the regions where the field is nonzero.

To treat large fields, we note that
for any given ${{\bf {r}}_{1}}$ we can locally
impose the optimal condition that the field perturbation vanishes by
introducing a {\em shifted} chemical potential
\begin{equation}
\mu ^{{\bf {r}}_{1}}\equiv \mu ^{B}-\phi ({{\bf {r}}_{1}}),\,  \label{mur1}
\end{equation}
and a {\em shifted} external field
\begin{equation}
\phi ^{{\bf {r}}_{1}}({{\bf {r}}})\equiv \phi ({{\bf {r}}})-\phi ({{\bf {r}}%
_{1}}).  \label{phir1}
\end{equation}
Since there is an arbitrary zero of energy and a constant external field
acts like a shift of the chemical potential
in the grand ensemble, we make no physical changes if we shift both
functions by the same amount. In particular $\rho ({\bf r};\ \mu
^{B},[\phi ])=\rho ({\bf r};\ \mu ^{{\bf {r}}_{1}},[\phi ^{{\bf {r}}
_{1}}]).$

The superscript ${\bf r}_{1}$ in $\mu ^{{\bf {r}}_{1}}$ indicates a particular
value of the chemical potential, which from Eq.\ (\ref{mur1})
depends parametrically on ${\bf r}_{1}$ through the local value of the field.
When $\phi ({{\bf {r}}_{1}})$ vanishes, then $\mu ^{{\bf {r}}_{1}}$
reduces to $\mu ^{B}.$ We define $\rho ^{{\bf r}_{1}},$
the {\em hydrostatic density}, by
\begin{equation}
\rho ^{{\bf r}_{1}}\equiv \rho ({\bf r};\ \mu ^{{\bf {r}}_{1}},[0])=\rho
(\mu ^{{\bf {r}}_{1}}).  \label{hydrorhodef}
\end{equation}
Thus $\rho ^{{\bf r}_{1}}$ is the density of the {\em uniform} fluid in zero
field at the shifted chemical potential $\mu ^{{\bf {r}}_{1}}$; equivalently
$\rho ^{{\bf r}_{1}}$ satisfies
\begin{equation}
\mu (\rho ^{{\bf r}_{1}})=\mu ^{{\bf {r}}_{1}}=\mu ^{B}-\phi ({{\bf {r}}_{1}}).
\label{hydromudef}
\end{equation}

Thus far, we have have merely introduced an equivalent (and apparently more
complicated!) way of describing the system in terms of a shifted field and
a shifted chemical potential. However this perspective immediately suggests
a very simple first approximation to the density response to a {\em slowly
varying} external field. Since $\phi ^{{\bf {r}}_{1}}({{\bf {r}}})$ by
construction vanishes for ${\bf r}={\bf r}_{1}$, if $\phi ^{{\bf {r}}_{1}}%
({{\bf {r}}})$ is sufficiently slowly varying, then the region
around ${\bf r}_{1}$ within a correlation length is essentially in zero
field. In that case the uniform hydrostatic density $\rho ^{{\bf r}_{1}}$
is clearly a good approximation to $\rho ({\bf r}_{1})$, the exact induced
density at ${\bf r}_{1}$. Moreover, when the field is more rapidly varying,
it is natural to introduce a second and even more accurate approximation
to the density response.

The hydrostatic density $\rho ^{{\bf r}_{1}}$ takes account only of the
local value of the field at ${\bf r}_1$ by a shift of the chemical
potential. The HLR equation improves on this ``local field'' approximation
by using {\em linear response theory} to determine the density change 
from the hydrostatic density induced by nonlocal values of the
shifted field $\phi ^{{\bf {r}}_{1}}({{\bf {r}}})$. Thus starting from
the uniform density $\rho ^{{\bf r}_{1}}$, we assume a
linear response to the shifted field, replacing
$\chi ^{-1}({\bf r}_{1},{\bf r}_{2};[\rho ])$ by
$\chi ^{-1}(r_{12};\rho ^{{\bf r}_{1}})$ in Eq.\ (\ref{linresponse}) and
setting $\delta \phi ({{\bf {r}}})=\phi ^{{\bf {r}}_{1}}({{\bf {r}}})$ and
$\delta \rho ({\bf r}_{2})=\rho ({\bf r} _{2})-\rho ^{{\bf r}_{1}}$.
Then the left side of Eq.\ (\ref{linresponse}) vanishes (giving the optimal
linear response condition), and we have
\begin{equation}
0=\int \!d{\bf r}_{2}\,\chi ^{-1}(r_{12};\rho ^{{\bf r}_{1}})[\rho ({\bf r}
_{2})-\rho ^{{\bf r}_{1}}],  \label{chilinhydro}
\end{equation}
which can be rewritten exactly using Eq.\ (\ref{chiinverse}) as
\begin{equation}
\rho ({\bf r}_{1})=\rho ^{{\bf r}_{1}}+\rho ^{{\bf r}_{1}}\int \!d{\bf r}%
_{2\,}c(r_{12};\rho ^{{\bf r}_{1}})[\rho ({\bf r}_{2})-\rho ^{{\bf r}_{1}}].
\label{HLR}
\end{equation}

This is our final result, which we refer to as the HLR equation. We view
this as an integral equation relating the hydrostatic density
$\rho ^{{\bf r}_{1}}$ to the full density
$\rho ({\bf r})$ and solve it self-consistently for all ${\bf r}_{1}$.
When $\phi ({\bf r}_{1})$ is known, we can immediately determine 
$\rho ^{{\bf r}_{1}}$ at each ${\bf r}_{1}$
from the local relation in Eq.\ (\ref{hydromudef}),
and then solve Eq.\ (\ref{HLR}) by iteration for all ${\bf r}_{1}$
to determine the full density response $\rho ({\bf r})$.
Conversely, for a given equilibrium density distribution $\rho ({\bf r})$
we can use HLR equation to determine the associated field $\phi ({\bf r})$.
This inverse solution of Eq.\ (\ref{HLR}) is particularly easy to carry
out, since we can determine the local field at each ${\bf r}_{1}$
separately, without iteration. Accurate results have been obtained for
many test cases with strong repulsive or attractive fields
\cite{hlr,katsov.weeks:incorporating}.

This requires in particular expressions for $\mu (\rho )$ and for the direct
correlation function $c(r_{12};\rho )$ of the {\em uniform} hard sphere fluid.
In this paper we will use the Percus-Yevick (PY) \cite{py} approximation
for $c(r_{12};\rho )$. This same function also arises from a self-consistent
solution of the HLR equation, where the density response to a hard core
field with $R_{v}=\sigma $ (equivalent to fixing a solvent particle at the
origin) is related to the uniform fluid pair correlation function. Thus this
self-consistent use of the HLR equation provides a physically suggestive way
of deriving the PY result for $c(r_{12};\rho )$ \cite{hlr}.
The PY $c(r_{12};\rho )$ has a very simple
analytical form and proves sufficiently accurate for our purposes here. Even
better results can be found if one uses the very accurate expressions for
the bulk $c(r_{12};\rho )$ and $\mu (\rho )$ as given by the GMSA
theory \cite{gmsa} as inputs to the HLR equation.

\subsection{Relation to the PY approximation for a hard core solute}

A spherical cavity acts like a hard core external field $\phi $ that
excludes the centers of all solvent molecules from the cavity region. We
take the center of the cavity as the origin of our coordinate system, so that
all distances are measured relative to the cavity center. Note that both the
hydrostatic density $\rho ^{{\bf {r}}_{1}}$
from Eq.\ (\ref{hydrorhodef}) and the full
density response $\rho ({\bf r}_{1})$ from Eq.\ (\ref{HLR}) vanish whenever
${\bf {r}}_{1}$ is located in the cavity. This exact ``hard core condition''
comes out naturally from the theory, and does not have to be imposed by hand
as in the GFM or the GMSA approaches.

To make contact with the PY approximation, recall that the {\em %
cavity-solvent direct correlation function} $C({\bf r}_{1};\rho ^{B},R_{v})$
for this system exactly satisfies
\begin{equation}
C({\bf r}_{1};\rho ^{B},R_{v})=\int d{\bf {r}}_{2}\chi ^{-1}(r_{12};\rho
^{B})[\rho ({\bf {r}}_{2})-\rho ^{B}].  \label{C21def}
\end{equation}
Thus $C({\bf r}_{1})$ is the function that replaces $-\beta \delta \phi (%
{\bf r}_{1})$ so that the linear response equation (\ref{linresponse}) gives
exact results when the
full density change relative to the bulk is used on the right hand side.
When this is compared to the HLR equation (\ref{chilinhydro}) for ${\bf r}
_{1}$ {\em outside} the cavity region (where $\rho ^{{\bf r}_{1}}=$ $\rho
^{B}$ and $\phi =0$) we see that the HLR equation predicts that $C({\bf r}%
_{1})$ vanishes. Thus for the HLR equation $\rho ({\bf r}_{1})$
vanishes inside the cavity region and $C({\bf r}_{1})$ vanishes outside.
This is the same as the PY approximation for the hard core 
cavity-solvent system \cite{hlr,jdwannrev}.

If $R_{v}$ is greater than $\sigma /2$, with $\sigma $ the solvent hard core
diameter, then an equivalent exclusion is achieved by replacing the hard
core external field by a hard core solute particle with (additive) diameter
\begin{equation}
\sigma _{v}\equiv 2R_{v}-\sigma . \label{sigmavdef}
\end{equation}
 From this it follows that if the PY
approximation for the bulk $c$ or $\chi ^{-1}$ is used, the density $\rho (%
{\bf r})$ predicted by the HLR equation is identical to that given by the PY
equation for the solute-solvent pair correlation function for a binary hard
sphere mixture in the limit that the concentration of the solute species
goes to zero \cite{HAB}. Since an exact analytical solution of the PY equation
for a binary HS mixture at arbitrary concentrations is known \cite
{py-mixture-solution}, we can take advantage of these results when computing
the excess grand free energy.

This equality of solutions of the HLR equation and the PY mixture equation
holds only for hard core cavity fields with radius $R_{v}\geq $ $\sigma /2$
or $\sigma _{v}\geq 0.$ As discussed below in Sec.\ \ref{small cavity}, for
{\em tiny cavities} with $R_{v}\leq $ $\sigma /2$ the HLR equation can be
solved directly and gives {\em exact} results for the density response if
exact bulk correlation functions are used, and very accurate results when
the PY approximation for the bulk $c$ is used. However, the corresponding PY
mixture solutions in this range of $R_{v}$ (arrived at formally by
taking $\sigma _{v}$ in Eq.\ (\ref{sigmavdef}) to be {\em negative})
are much less accurate. This inaccuracy arises from using the PY mixture
solutions for negative $\sigma _{v}$.
The {\em direct} solution of the PY cavity-solute equation for a tiny
cavity, where a given approximation for the bulk $c$ is used along with the
PY approximation that $C({\bf r}_{1})$ vanishes outside the cavity and $\rho
({\bf r}_{1})$ vanishes inside, gives the same accurate results as the HLR
equation. However, for more general external fields, the HLR equation is
quite distinct from the PY approximation, and is generally more accurate;
it has given good results for a
wide range of fields \cite{hlr,katsov.weeks:incorporating,jdwannrev}.
This additional flexibility of the
HLR equation will be required later in this paper when we discuss alternate
density routes to the free energy.

\section{Thermodynamic pathways to the free energy}

In this section we discuss three different exact thermodynamic pathways for
obtaining the excess free energy of inserting a cavity into a hard sphere
fluid. The first two are well known, and the third describes a new density
route that may have some computational advantages in other applications. We
use the HLR equation to provide the needed structural information in all
cases. We believe our calculation here represents the first use of a density
route to obtain the excess free energy for this system. We then describe a
simple but less accurate route to the free energy based on use of the
partition function for the GFM.

\subsection{Compressibility route}

In this route the excess free energy is determined by varying the chemical
potential of the system while the external field $\phi ({\bf r})$ producing
the cavity with radius $R_{v}$ remains constant. In the grand canonical
ensemble the average number of particles $\langle N\rangle $ is given by
\begin{equation}
\frac{\partial \Omega }{\partial \mu }=-\langle N\rangle , \label{compressibility}
\end{equation}
where $\Omega (\mu ,[\phi ])$ is the grand free energy. We can then
calculate the free energy difference between the final state of interest and
the trivial ideal gas state of zero density with $\mu =-\infty $ and $\Omega
=0$ by integration:
\begin{equation}
\Omega (\mu ^{B},[\phi ])=-\int_{-\infty }^{\mu ^{B}}d\mu \langle N\rangle \,=
-\int_{-\infty }^{\mu ^{B}}d\mu \int d{\bf {%
r}}\rho ({\bf {r;}}\mu ,[\phi ]).  \label{omegaNmu}
\end{equation}
Then $\Delta \Omega _{v}\equiv \Omega (\mu ^{B},[\phi ])-\Omega (\mu
^{B},[0]),$ the desired excess grand free energy of the nonuniform fluid
relative to the uniform bulk state, is given by
\begin{equation}
\Delta \Omega _{v}=-\int_{-\infty }^{\mu ^{B}}d\mu \int d{\bf {r}}_{1}\{\rho
({\bf {r}}_{1}{\bf {;}}\mu ,[\phi ])-\rho (\mu )\}.  \label{delomegaNmu}
\end{equation}
As before, $\rho (\mu )$ gives the density of the uniform hard sphere
solvent fluid as a function of the chemical potential.

Since $\rho ({\bf {r;}}\mu ,[\phi ])$ vanishes inside the cavity,
Eq.\ (\ref {delomegaNmu}) shows there is a term in the excess free
energy proportional to the cavity volume $v$ given by
\begin{equation}
v\int_{-\infty }^{\mu ^{B}}d\mu \rho (\mu )=v\int_{0}^{\rho ^{B}}d\rho \frac{%
d\mu }{d\rho }\rho =vp^{B},  \label{wallsumrule}
\end{equation}
on using the thermodynamic relation $\rho (\partial \mu /\partial \rho %
)_{T}=(\partial p/\partial \rho )_{T}.$ This exact leading order term
for large $v$ is determined when using the compressibility route so that
$p^{B}$ is the uniform fluid pressure calculated by the compressibility
route \cite{hendersonSPT}.

The term in curly brackets in Eq.\ (\ref {delomegaNmu}) can be rewritten
in a more convenient form for calculations by using the inverse relation
to Eq.\ (\ref{C21def}) for a general chemical potential $\mu $:
\begin{equation}
\rho ({\bf {r}}_{1}{\bf {;}}\mu ,[\phi ])-\rho (\mu )=\int d{\bf {r}}%
_{2}\chi (r_{12};\mu )C({\bf r}_{2};\rho (\mu ),R_{v}),
\label{c21 linear response}
\end{equation}
where $\chi (r_{12};\mu )\equiv $ $\rho \delta ({\bf r}_{1}-{\bf r}%
_{2})+\rho ^{2}[g(r_{12})-1]$ is the usual linear response function of the
uniform solvent fluid and $g(r)$ is the radial distribution function.
Substituting into Eq.\ (\ref{delomegaNmu}) and carrying out the integration
over ${\bf r}_{1},$ we have the formally exact result \cite{pdt}
\begin{eqnarray}
\beta \Delta \Omega _{v} &=&-\beta \int_{-\infty }^{\mu ^{B}}d\mu \hat{\chi}%
(0,\mu )\int d{\bf {r}}_{2}C({\bf r}_{2};\rho (\mu ),R_{v})  \nonumber \\
&=&-\int_{0}^{\rho ^{B}}d\rho \,\,\hat{C}(0;\rho ,R_{v}).
\label{delomegachat}
\end{eqnarray}
Here $\hat{\chi}(0,\mu )$ is the $k=0$ value of the Fourier transform of $%
\chi $, with a similar definition for $\hat{C}(0;\rho ,R_{v})$. In the last
equality we used the uniform fluid compressibility relation $\beta \hat{\chi}%
(0,\mu )=d\rho (\mu )/d\mu $ to change variables to an integration over
density. We will explicitly solve the HLR equation for $R_{v}\leq \sigma /2$
in Sec.\ \ref{small cavity} below, and from the equivalence between the HLR
equation and PY mixture equation for $R_{v}\geq \sigma /2$, we can use the
exact solution of the PY mixture equation to obtain $\hat{C}$ at larger
$R_{v}$. Thus we can analytically carry out the integration in
Eq.\ (\ref{delomegachat}) for all $R_{v}$.

\subsection{Virial route}

We now consider a different thermodynamic pathway, which was first used in
scaled particle theory \cite{spt0,spt}. Here we keep the chemical potential
fixed at $\mu ^{B}$ and vary the {\em range} of the external hard core
field by a scaling parameter $\lambda $,
defining $\phi _{\lambda }({\bf {r}})\equiv \phi ({\bf{r}}/\lambda )$.
For the hard core cavity field of interest here, as
$\lambda $ is varied from $0$ to $1$ the radius of the exclusion zone then
varies from $0$ to $R_{v}$. Since the density is generally related to the
external field in the grand ensemble by
\begin{equation}
\frac{\delta \Omega }{\delta \phi ({\bf {r}})}=\rho ({\bf {r}}),
\label{virial}
\end{equation}
the desired free energy difference is given by integration:
\begin{equation}
\beta \Delta \Omega _{v}=\int_{0}^{1}d\lambda \int d{\bf {r}}\rho _{\lambda
}({\bf {r}})\frac{\partial \beta \phi _{\lambda }({\bf {r}})}{\partial
\lambda }.  \label{mu virial formal}
\end{equation}
Here $\rho _{\lambda }({\bf {r}})\equiv \rho ({\bf {r;}}\mu ^{B},[\phi
_{\lambda }]).$ This formula is quite general and holds for any
$\lambda $-dependent potential that vanishes for $\lambda =0.$
By exploiting special
properties of the scaled hard core potential (the derivative of the
Boltzmann factor of a hard core potential is a delta function)\ it is easy
to show that Eq.\ (\ref{mu virial formal}) can be exactly rewritten as
\begin{equation}
\beta \Delta \Omega _{v}=4\pi R_{v}^{3}\int_{0}^{1}d\lambda \lambda ^{2}\rho
_{\lambda }(\lambda R_{v}).  \label{mu virial final}
\end{equation}
Here $\rho _{\lambda }(\lambda R_{v})$ is the {\em contact density} at the
surface of the scaled exclusion zone with radius $\lambda R_{v}$.
As in the compressibility route, we can analytically carry out the
integration in the virial route to obtain solvation free energies for
cavities for all $R_{v}$. The equivalent PY solution for binary hard sphere
mixtures is used for the contact densities for all $\lambda R_v$'s larger
than $\sigma/2$, while the explicit solution of the HLR equation is used
for the $\lambda R_v$'s smaller than $\sigma/2$.

\subsection{Density routes}

\label{subsec density route} In addition to these particular pathways, we
can also imagine {\em directly changing} the equilibrium density from $\rho
^{B}$ to $\rho ({\bf {r}})$ over some convenient pathway specified by a
coupling parameter $\lambda ,$ while taking account of the associated
changes in $\Omega $ and $\phi ({\bf {r}})$. Integrating Eq.\ (\ref{mu virial
formal}) by parts to make $\rho _{\lambda }({\bf {r}})$ explicitly the
controlling variable, we have exactly
\begin{equation}
\beta \Delta \Omega _{v}=\int d{\bf {r}}\rho ({\bf {r}})\phi ({\bf {r}}%
)-\int_{0}^{1}d\lambda \int d{\bf {r}}\,\phi _{\lambda }({\bf {r}})\frac{%
\partial \rho _{\lambda }({\bf {r}})}{\partial \lambda }.
\label{density route 0}
\end{equation}
Here $\phi _{\lambda }({\bf {r}})$ is the external field consistent with the
specified density profile, so that $\rho ({\bf r};\mu ^{B},[\phi _{\lambda
}])=\rho _{\lambda }({\bf {r}}).$ For a given density field
$\rho _{\lambda }({\bf {r}})$,
the HLR equation (\ref{HLR}) can be solved inversely to obtain the
associated hydrostatic density field $\rho _{\lambda }^{{\bf {r}}}$. Using
Eq.\ (\ref{hydromudef}), $\phi _{\lambda }({\bf {r}})$ at each ${\bf r}$ is
locally related to $\rho _{\lambda }^{{\bf {r}}}$ through $\mu (\rho ).$
Here we used the accurate Carnahan-Starling expression 
\cite{carnahan.starling:equation}
for $\mu (\rho )$.

Most workers have considered a simple {\em linear} density path where
\begin{equation}
\rho _{\lambda }({\bf {r}})=\rho ^{B}+\lambda [\rho ({\bf {r}})-\rho ^{B}].
\label{linearpath}
\end{equation}
This has some theoretical advantages since $\partial \rho _{\lambda }({\bf {r%
}})/\partial \lambda $ is independent of $\lambda $ and has been
successfully used in numerical calculations of the surface tension of the
liquid-vapor interface \cite{katsov.weeks:incorporating}.
However, when $\phi ({\bf {r}})$ has a hard core (or
is strongly repulsive), then the region near $\lambda =1$ in the $\lambda $%
-integration in Eq.\ (\ref{density route 0}) must be treated carefully, since
for ${\bf {r}}$ in the hard core region $\partial \rho _{\lambda }({\bf {r}}%
)/\partial \lambda $ is constant, while $\phi _{\lambda }({\bf {r}})$ must
tend to infinity as $\lambda \to 1.$ Although the singularity in the
potential is integrable (for a hard core potential the divergent term in $%
\beta \phi _{\lambda }$ goes as $-\ln (1-\lambda )$ and could be treated
separately), in any case large contributions to the integral arise from a
relatively small interval near $\lambda =1$. This could cause problems in a
numerical integration.

To illustrate the computational advantages and flexibility that different
pathways can provide, we introduce here a new density route that removes
this possible difficulty. We consider a path that is linear in the {\em %
square root of the density}:
\begin{equation}
\rho _{\lambda }^{1/2}({\bf {r}})\equiv (\rho ^{B})^{1/2}+\lambda [\rho
^{1/2}({\bf {r}})-(\rho ^{B})^{1/2}],  \label{linsquareroot}
\end{equation}
where $\rho _{\lambda }^{1/2}({\bf {r}})=\sqrt{\rho _{\lambda }({\bf {r}})}$
, etc. For this pathway we have
\begin{equation}
\frac{\partial \rho _{\lambda }({\bf {r}})}{\partial \lambda }=2\rho
_{\lambda }^{1/2}({\bf {r}})\frac{\partial \rho _{\lambda }^{1/2}({\bf {r}})%
}{\partial \lambda }.  \label{drhodlambdasqroot}
\end{equation}
Both factors on the right side of Eq.\ (\ref{drhodlambdasqroot}) are easy to
determine from Eq.\ (\ref{linsquareroot}). The numerical integration in Eq.\ (%
\ref{density route 0}) can now be carried out straightforwardly since the $%
\rho _{\lambda }^{1/2}({\bf {r}})$ factor in Eq.\ (\ref{drhodlambdasqroot})
will cause $\partial \rho _{\lambda }({\bf {r}})/\partial \lambda $ to tend
to zero exponentially fast wherever $\phi _{\lambda }({\bf {r}})$ becomes
large. Results using this path are reported below. Other paths implementing
this idea exist and we have not tried to make an optimal choice.

\subsection{Gaussian Field Model}

Finally we consider an alternative approach, the Gaussian field model (GFM)
\cite{gft}, that for hard core fields has many common elements
with the HLR method. The GFM describes density fluctuations in a uniform
fluid with average density $\rho $ by an effective quadratic Hamiltonian
\begin{equation}
{\cal H}^{B}=\frac{k_{B}T}{2}\int d{\bf {r}}_{1}\int d{\bf {r}}_{2}\delta
\check{\rho}({\bf {r}}_{1})\chi ^{-1}({\bf {r}}_{12};\rho )\delta
\check{\rho}({\bf {r}}_{2}),  \label{gft hamil}
\end{equation}
where $\delta \check{\rho}({\bf {r)=}}$ $\check{\rho}({\bf {r)-}}\rho $ with
$\check{\rho}({\bf {r)}}$ the microscopic density. The partition function for
a system in an external field ${\phi ({\bf {r}})=\phi }_{0}{({\bf {r}})+\phi }%
_{1}{({\bf {r}}),}$ with ${\phi }_{0}$ a hard core field producing a cavity
of radius $R_{v}$ and ${\phi }_{1}$ a weaker perturbation, is then assumed
to be given by \cite{gft}

\begin{eqnarray}
\Xi _{v} &=&\int {\cal D}\check{\rho}({\bf {r}})\;\left\{ \prod_{{\bf {r}}\in
v}\delta [\check{\rho}({\bf {r}})]\right\}  \nonumber \\
&&\times \exp [{-\beta {\cal H}}^{B}{+\beta \int d}r\check{\rho}({\bf 
r})\phi_1 (%
{\bf r})].  \label{gft par formal}
\end{eqnarray}
The product of delta functions imposes the constraint that the density
vanish inside the cavity. Inserting a Fourier representation for the $\delta
$-functions and formally integrating $\check{\rho}({\bf {r}})$ from
$-\infty $ to $\infty$ yields a Gaussian approximation for the partition
function, as discussed below.

Moreover, using the same approximations, by functionally differentiating $%
\Xi _{v}$ with respect to the field, one obtains the nonuniform singlet
density in the GFM. In the case of a pure hard core field with ${\phi }_{1}=0,$
the density response to a cavity with radius $R_{v}$ is given by
\begin{equation}
\rho ({\bf {r}}_{1})=\rho ^{B}-\rho ^{B}\int_{v}d{\bf {r}}_{2}\int_{v}d{\bf {%
r}}_{3}\chi (r_{12};\mu ^{B})\chi _{in}^{-1}({\bf {r}}_{2},{\bf {r}}_{3}).
\label{gft rho}
\end{equation}
The integrations are restricted to the cavity region, as
indicated by the subscript $v$ on the integral symbols. Here $\chi _{in}^{-1}
$ is the inverse of the restricted linear response function $\chi
_{in}(r_{12};\mu ^{B})$, which equals $\chi (r_{12};\mu ^{B})$ if both ${\bf
{r}}_{1}$ and ${\bf {r}}_{2}$ are in the cavity region and equals zero
otherwise. Thus $\chi _{in}^{-1}$ is nonzero only inside the cavity and
satisfies
\begin{equation}
\int_{v}d{\bf {r}}_{2}\chi (r_{12};\mu ^{B})\chi _{in}^{-1}({\bf {r}}_{2},%
{\bf {r}}_{3})=\delta ({\bf {r}}_{1}-{\bf {r}}_{3}),  \label{chiindef}
\end{equation}
when both ${\bf {r}}_{1}$ and ${\bf {r}}_{3}$ are in the cavity region.
Comparing Eq.\ (\ref{gft rho}) to Eq.\ (\ref{c21 linear response}), one can
identify the cavity-solvent direct correlation function in the GFM as
\begin{equation}
C({\bf {r}}_{2})=-\rho ^{B}\int_{v}d{\bf {r}}_{3}\chi _{in}^{-1}({\bf {r}}%
_{2},{\bf {r}}_{3}).  \label{gft c21}
\end{equation}

By properties of $\chi _{in}^{-1},$ the GFM $C({\bf {r}})$ vanishes outside
the cavity region and $\rho ({\bf {r}})$ in Eq.\ (\ref{gft rho}) vanishes
inside. Thus the GFM gives exactly the same solution for the density
response to a hard core external field as the PY or the HLR equations. (In
the more general case where there is an additional perturbation potential ${%
\phi }_{1}$, the various approaches differ. The GFM can be shown to treat
the softer tail using the mean spherical approximation, which is different
from and generally less accurate than the hydrostatic shift used in the HLR
equation.)

Thus for cavities or hard core solutes all structurally based routes to the
excess free energy will give the same results when using the GFM or the HLR
equation. In addition, the GFM partition function also provides a direct and
very simple route to the free energy \cite{gft}. However this route is
inherently approximate because Eq.\ (\ref{gft par formal})
is not really a free energy functional for the whole configuration space,
but rather a restricted one describing only the space outside
of the specified cavity region. This functional may legitimately describe
subsequent {\em small} perturbations of $\phi_1$ outside of the cavity, but
it does not contain enough information about the functional dependence on
the cavity volume in the first place. Moreover the approximations made in
evaluating the GFM partition function do not build in the fact that in
grand canonical ensemble, the thermodynamic properties should depend on
$\mu-\phi$ rather than on $\mu$ and $\phi$ individually. Thus it is also not
consistent with the free energy prediction from the compressibility route,
which integrates over states at different chemical potentials but with a
fixed hard core always present.

Evaluating the Gaussian integrals in Eq.\ (\ref{gft par formal}),
the excess grand
free energy arising from a cavity with radius $R_{v}$ is given by
\begin{eqnarray}
\beta \Delta \Omega _{v} &=&-\log \Xi _{v}/\Xi ^{B}  \nonumber \\
&=&-\frac{1}{2}\rho ^{B}\hat{C}(0;\rho ^{B},R_{v})+\log {(\det {\chi _{in}})}.
\label{gft mu}
\end{eqnarray}
Here $\Xi _{v}$ denotes the partition function with no particles in the
cavity region and $\Xi ^{B}$ is the uniform bulk partition function. The
term involving $\hat{C}$ in Eq.\ (\ref{gft mu}) arises from the Gaussian
integration of $\check{\rho}$ in Eq.\ (\ref{gft par formal}) and Eq.\ (\ref{gft
c21}). When compared with the exact Eq.\ (\ref{delomegachat}) from the
compressibility route, we see the GFM effectively approximates $\hat{C}%
(0;\rho ,R_{v})$ for intermediate density values by $\left( \rho /\rho
^{B}\right) \hat{C}(0;\rho ^{B},R_{v})$. This free energy contribution has a
form similar to a harmonic oscillator with $\left( \rho /\rho ^{B}\right)
\hat{C}(0;\rho ^{B},R_{v})$ analogous to the restoring force.
The second term is a result of the reduction of the configuration space.

An alternate perspective considers the average probability $P_{v}(N)$ of
finding $N$ particles in the volume $v$ with radius $R_{v}$. The probability
of inserting a cavity is thus $P_{v}(0)$. Then a formally exact expression
for the excess free energy $\beta \Delta \Omega _{v}$ is
\begin{equation}
\beta \Delta \Omega _{v}=-\log P_{v}(0)=-\log \frac{\Xi _{v}[0]}{%
\sum_{N=0}^{N_{max}}\Xi _{v}[N]}.  \label{gft pn}
\end{equation}
Here $\Xi _{v}[N]$ is the constrained partition function when $N$ particles
are in the specified volume. This formula has been successfully used in the
information theory approach developed by Hummer, Pratt and
coworkers \cite{hummer.garde.ea:information}.

If the GFM is used to approximate the partition
functions in Eq.\ (\ref{gft pn}) by replacing the product of $\delta $%
-functions in Eq.\ (\ref{gft par formal}) by the single average constraint
$\delta [\int_{v}d{\bf {r}}\hat{\rho}({\bf {r}})-N]$, one arrives at
a Gaussian approximation \cite{hummer.garde.ea:information,lcw} for
$\Xi _{v}[N].$ This ``discrete'' approximation for $\beta \Delta \Omega _{v}$
based on this use of the GFM is
\begin{equation}
\beta \Delta \Omega _{v}=-\log {\frac{e^{-\bar{N}^{2}/2\chi _{v}}}{%
\sum_{N=0}^{N_{max}}e^{-(N-\bar{N})^{2}/2\chi _{v}}}}  \label{gft mu dis}
\end{equation}
where
\begin{equation}
\chi _{v}\equiv \int_{v}d{\bf {r}}_{1}\int_{v}d{\bf {r}}_{2}\chi (|{\bf {r}}%
_{1}-{\bf {r}}_{2}|),  \label{chiv}
\end{equation}
and $\bar{N}=\rho ^{B}v$.

We determined the solvation free energy $\beta \Delta \Omega _{v}$ for the
GFM using both the continuum version, Eq.\ (\ref{gft mu}), and the discrete
version, Eq.\ (\ref{gft mu dis}). For the uniform $\chi $
we used the PY result.
To estimate $\log \det \chi _{in}$ we expanded $\chi _{in}$ in the volume $v$
using two orthogonal basis functions. A single constant basis function was
used in Ref. \cite{lcw}. This is exact for tiny cavities with $R_{v}\leq \sigma /2$, as
can be seen using Eq.\ (\ref{chiin small}) below. We chose one basis function
to be constant. The other was taken to be $j_{0}(R_{v}r/\pi )$, the
zeroth order spherical Bessel function with its first node fixed at $r=R_{v}$,
but made orthogonal to the first (constant) basis function. The second
basis function is thus a linear combination of $j_{0}(R_{v}r/\pi )$ 
and a constant. This was introduced to test the accuracy of the one basis
function approximation previously used and hopefully will give improved results
for larger $R_{v}$.

\section{Results for larger cavities with $R_{v}>\sigma /2$}

\begin{figure}[t]
\includegraphics[width=80mm,height=60mm]{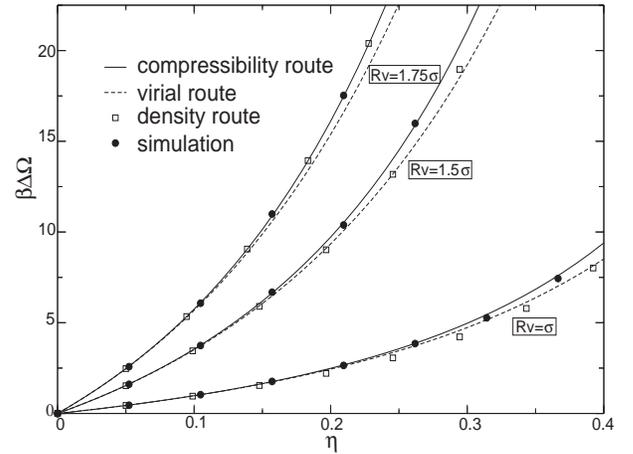}
     \caption{The excess free energy predicted by the three thermodynamic routes
     for cavity radii $R_{v}=\sigma$, 1.5$\sigma$ and 1.75$\sigma$,
     compared with simulation data. $\eta$ is
     the packing fraction and is equal to $\pi\rho^B\sigma^3/6$.}
     \label{fig:large}
\end{figure}

We now discuss the solvation free energies given by the various
pathways for a cavity with $R_{v}>\sigma /2$, equivalent to a physically
realizable hard core solute particle with diameter $\sigma _{v}>0.$ (Results
for tiny cavities with $R_{v}\leq \sigma /2$ are discussed in Sec.\ \ref
{small cavity} below.) We use the simplest version of the theory, where the
PY approximation is used for the uniform fluid correlation functions.
Fig.\ (\ref{fig:large}) give the solvation free energy
$\beta \Delta \Omega _{v}$ from the
different pathways as a function of the packing fraction $\eta =\pi \rho
^{B}\sigma ^{3}/6$ for $R_{v}/\sigma =1,\,1.5,\,1.75,$ where the results can
be compared to computer simulations  \cite{gaus-sim} of Crooks and Chandler.
Note that the
volume of a spherical cavity with $R_{v}=1.75\sigma $ is over 42 times greater
than that of a solvent particle. For $R_{v}=\sigma $ the results also give the
excess chemical potential as a function of density for the uniform hard
sphere fluid.

As discussed above, we can obtain analytical expressions for $\Delta\Omega_v$
for both the compressibility and virial routes. The compressibility route gives
\begin{eqnarray}
\beta\Delta\Omega _v&=&\frac{\eta(-2+7\eta-11\eta^2)}{2(1-\eta)^3}-\log(1-\eta)+%
\frac{18\eta^3}{(1-\eta)^3}\frac{R_v}{\sigma}\nonumber\\
&&-\frac{18\eta^2(1+\eta)}{(1-\eta)^3}\frac{{R_v}^2}{\sigma^2}+
\frac{8\eta(1+\eta+\eta^2)}{(1-\eta)^3}\frac{{R_v}^3}{\sigma^3},
\end{eqnarray}
and the virial route gives
\begin{eqnarray}
\beta\Delta\Omega _v&=&\frac{\eta(-2+7\eta-5\eta^2)}{2(1-\eta)^3}
-\log(1-\eta)\nonumber\\
&&-\frac{18\eta^2(1-\eta)}{(1-\eta)^3}\frac{{R_v}^2}{\sigma^2}
+\frac{8\eta(1+\eta-2\eta^2)}{(1-\eta)^3}%
\frac{{R_v}^3}{\sigma^3}.~~~~~~ 
\end{eqnarray}
Results for the density route
and for the GFM are computed numerically.

\begin{figure}[t]
\includegraphics[width=80mm,height=60mm]{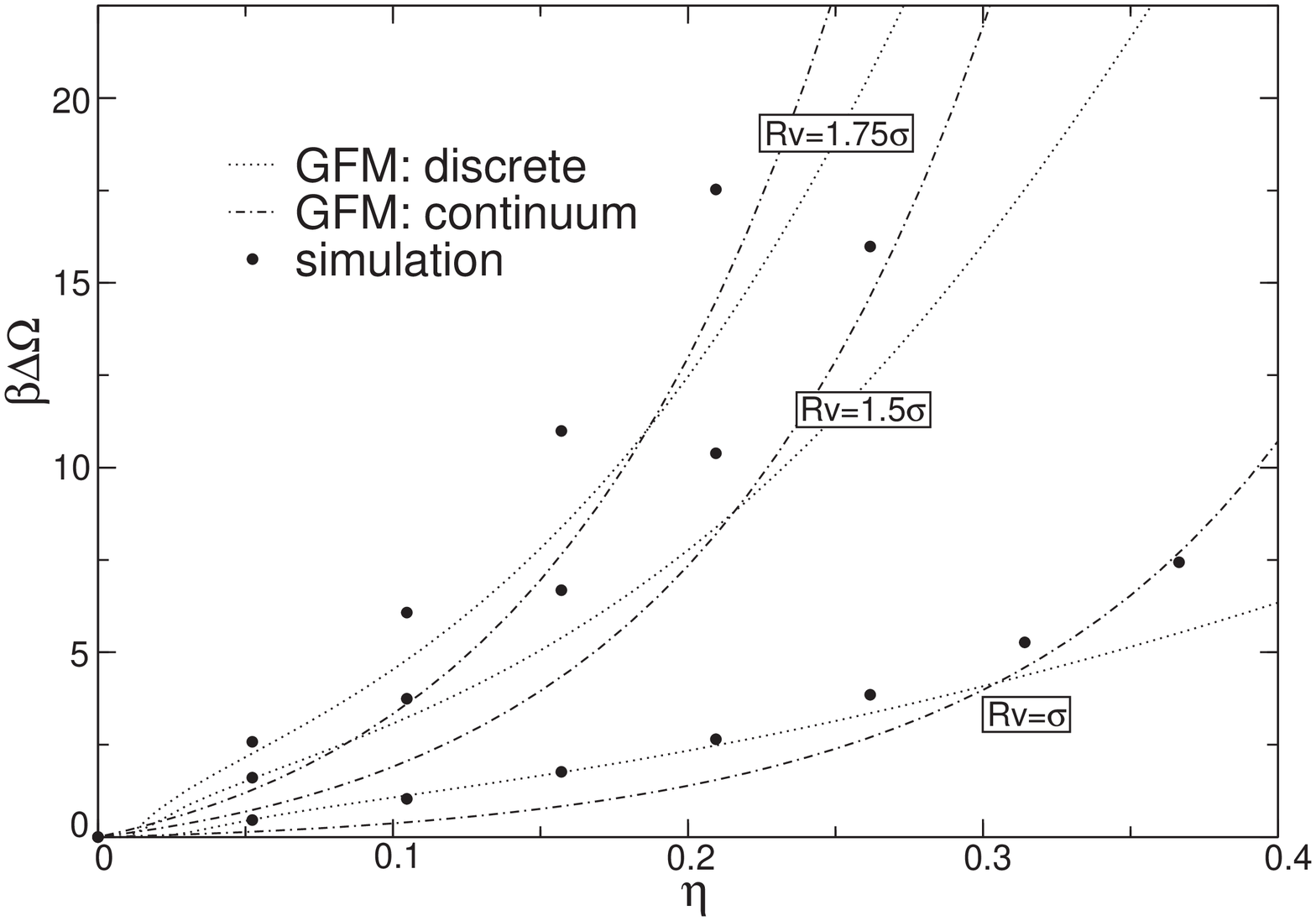}

    \caption{The excess free energy predictions by both the discrete and the
    continuum versions of the GFM, plotted for the
    same $R_{v}$ values and on the same scale as in Fig.\ \ref{fig:large}.}
     \label{fig:largegft}
\end{figure}

In this range of $R_{v}$ there is good agreement except at the highest
densities between the compressibility, virial and density routes, with
best results overall arising from the compressibility route. The direct
GFM predictions in Fig.\ (\ref{fig:largegft}) from the
partition function are less satisfactory. Both the discrete and the
continuum versions of the GFM give results that approach zero incorrectly as
$\rho ^{B}\to 0$, and the continuum values are consistently too large at
high density while the discrete values are too small. The discrete version
of the GFM uses a Gaussian approximation for the constrained partition
functions and gives less accurate results than could be obtained from a fit
to accurate values of $\langle N\rangle $ and $\langle N^{2}\rangle $ as
in the information theory approach \cite{hummer.garde.ea:information}.

\begin{figure}[t]
\includegraphics[width=80mm,height=60mm]{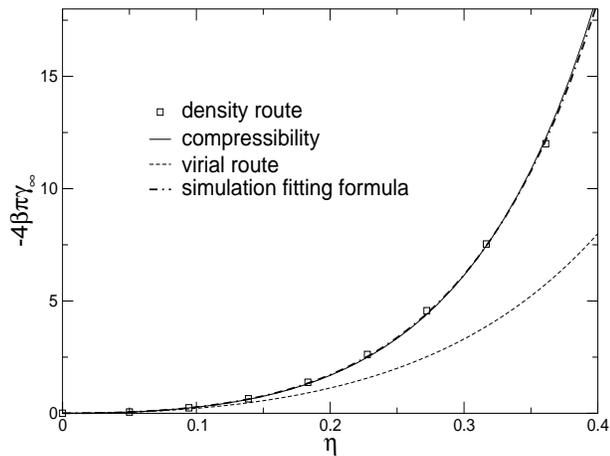}
       \caption{Surface tension predicted by the thermodynamic routes
compared with the simulation fitting formula.}
       \label{fig:surfacetension}
\end{figure}

As $R_{v}\to \infty $, the surface of the cavity approaches that of a planar
wall. As shown in Eq.\ (\ref{wallsumrule}) there is a diverging term in the
excess free energy given by the cavity volume $v=4\pi R_{v}^{3}/3$ times the
bulk pressure $p^{B}$, and the more interesting quantity to calculate is the
surface term $\gamma _{v}$, given by
\begin{equation}\label{gammadef}
\beta \gamma _{v}=\frac{\beta \Delta \Omega _{v}-\beta p^{B}v}{4\pi R_{v}^{2}}.
\label{gamma v}
\end{equation}
The surface tension of the planar wall is then $\gamma _{\infty }$. In the
present case, both the compressibility and the virial routes give analytical
expressions for $\beta \Delta \Omega _{v}$ which depend on $R_{v}$ as
a polynomial:
$a_{0}+a_{1}R_v/\sigma+a_{2}R_{v}^{2}/\sigma^2+a_{3}R_{v}^{3}/\sigma^3$.
The coefficient $a_{3}$ thus gives another route to the bulk pressure on
taking the wall limit. The $p^B$ used in Eq.\ (\ref{gammadef}) has to
agree with the prediction from the $a_3$ so that $\gamma _\infty$ is
finite. As discussed earlier, the compressibility $\beta \Delta \Omega _v$
yields the same bulk pressure as given by the accurate uniform fluid PY
compressibility equation of state. The virial route does not automatically
build in this consistency, and the pressure predicted from the coefficient
$a_{3}$ is less accurate than the uniform fluid PY virial equation of state.

The coefficient of the quadratic term then gives the surface tension 
$\beta\gamma _{\infty }=a_{2}/4\pi \sigma^2$.
Using the compressibility route we find
\begin{equation}
-4\pi \beta \gamma _{\infty }\sigma^2=\frac{18\eta ^{2}(1+\eta )}{(1-\eta )^{3}},
\label{gamma infty comp}
\end{equation}
while the virial route gives
\begin{equation}
-4\pi \beta \gamma _{\infty }\sigma^2=\frac{18\eta ^{2}(1-\eta )}{(1-\eta )^{3}}.
\label{eq:gamma infty virial}
\end{equation}
The $\gamma _{\infty }$ obtained by the compressibility route coincides with
that given by scaled particle theory \cite{spt0, spt, j-henderson-simulation}.
We obtained $\gamma _{\infty }$ numerically for the density route.

These results can be compared to the quasi-exact formula \cite
{d-henderson-formula}
\begin{equation}
-4\pi \beta \gamma _{\infty }\sigma^2=\frac{18\eta ^{2}(1+\frac{44}{35}\eta -\frac{4%
}{5}\eta ^{2})}{(1-\eta )^{3}},  \label{exact gamma}
\end{equation}
which fits simulation data \cite{j-henderson-simulation} and imposes the
known first and second surface virial coefficients \cite{bellemans}. As
shown in Fig.\ (\ref{fig:surfacetension}) the compressibility and density routes
give excellent results, while the virial route is much less satisfactory.

This can be understood since the virial route uses only the contact
densities at the fixed bulk density. The HLR equation is least accurate
for the contact density at high bulk density and for large $R_v$,
while the density response away from the solute is more accurate. On the
other hand, the compressibility and density routes make use of the density
response at all distances and over a range of densities from low density to
the final $\rho ({\bf {r}})$ where the HLR equation is more accurate.

\section{A special regime: tiny cavities}
\label{small cavity}

\subsection{Exact results}

The density response to a {\em tiny cavity} with $R_{v}\leq \sigma /2$ is
especially simple, since the center of at most one hard core solvent
particle can lie anywhere within such a region \cite{spt0, spt}. This 
fact allows one to
determine {\em exactly} both the density response to a tiny cavity and the
solvation free energy. Here we will compare these exact results to the
predictions of the HLR and GFM approaches

As in Eq.\ (\ref{gft pn}), the solvation free energy is directly related to
the average probability that no particle are in the cavity region
\begin{equation}
e^{-\beta \Delta \Omega _{v}}=P_{v}(0),  \label{prob of cavity}
\end{equation}
where $P_{v}(N)$ is the probability of finding $N$ particles simultaneously
in the region with volume $v=4\pi R_{v}^{3}/3$. When $R_{v}\leq \sigma /2$ ,
the region can hold no more than one solvent particle, so that
\begin{equation}
P_{v}(0)+P_{v}(1)=1,  \label{norm1}
\end{equation}
and
\begin{equation}
\rho ^{B}v=\,\langle N\rangle _{v}=P_{v}(1).  \label{average n}
\end{equation}
Substituting these into Eq.\ (\ref{prob of cavity}) we thus find the {\em %
exact} result \cite{spt0,spt}
\begin{equation}
\beta \Delta \Omega _{v}=-\log (1-\rho ^{B}v).  \label{exact mu}
\end{equation}

This argument can be extended to show that the {\em exact} density response
to a tiny cavity is \cite{ReissC74}:
\begin{equation}
\rho ({\bf {r}})=\frac{\rho ^{B}[1-\int_{v}d{\bf {r}}^{\prime }\rho ^{B}g(|%
{\bf {r}}^{\prime }-{\bf {r}}|)]}{(1-\rho ^{B}v)}  \label{exactrhotinycavity}
\end{equation}
for ${\bf {r}}$ outside $v$, with $\rho ({\bf {r}})=0$ for ${\bf {r}}$
inside. Here $g(r)$ is the exact radial distribution function for the
uniform solvent fluid. Note that the contact density $\rho (R_{v})=\rho
^{B}/(1-\rho ^{B}v)$ is exactly determined independent of the details of $%
g(r)$, since the corresponding $g(|{\bf {r}}^{\prime }-{\bf {r}}|)$ in Eq.\ (%
\ref{exactrhotinycavity}) vanishes for all ${\bf {r}}^{\prime }$ inside $v$.
This result is valid as long as the inserted region $v$ can hold no more
than one solvent particle, so Eq.\ (\ref{exactrhotinycavity}) also holds for
solvents with a hard core pair potential plus a softer tail.

\begin{figure}[t]
     \includegraphics[width=80mm,height=60mm]{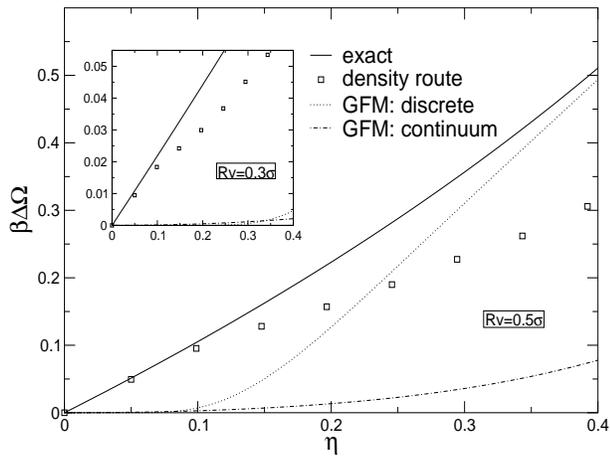}
\caption{The compressibility and virial routes are exact for the tiny cavity
regime. The density route is plotted along with the GFM results to compare
with the exact free energy predictions for $R_v = 0.3\sigma$ and $0.5\sigma$.}
\label{fig:rvsmall}
\end{figure}

\subsection{Structural predictions of the GFM and HLR methods}

Now let us examine the GFM result in Eq.\ (\ref{gft rho}) for the case of a
density response to a tiny cavity. Since $g(|{\bf {r}}_{1}-{\bf {r}}_{2}|)=0$
when both ${\bf {r}}_{1}$ and ${\bf {r}}_{2}$ are in $v$, then $\chi _{in}(%
{\bf {r}}_{1},{\bf {r}}_{2})=\rho ^{B}\delta ({\bf {r}}_{1}-{\bf {r}}_{2})-({%
\rho ^{B})}^{2}$ inside $v$. It is easy to see from Eq.\ (\ref{chiindef})
that the inverse function $\chi _{in}^{-1}$ then has the simple form 
\cite{katsov:point}
\begin{equation}
\chi _{in}^{-1}({\bf {r}}_{1},{\bf {r}}_{2})=\frac{\delta ({\bf {r}}_{1}-%
{\bf {r}}_{2})}{\rho ^{B}}+\frac{1}{1-\rho ^{B}v}.  \label{chiin small}
\end{equation}
When Eq.\ (\ref{chiin small}) is inserted into Eq.\ (\ref{gft rho}) to obtain
the GFM density response to a tiny cavity, we recover the {\em exact}
expression for $\rho ({\bf {r}})$ given in Eq.\ (\ref{exactrhotinycavity}),
provided that the exact uniform fluid $g(r)$ or $c(r)$ is used.
If approximate (say PY) results are used to
describe the uniform fluid response functions then strictly speaking the GFM
and HLR predictions for $\rho ({\bf {r}})$ will not be exact for all ${\bf {%
r.}}$ However, the contact density $\rho ({R}_{v})$ is exact in any case,
since, as noted earlier, this requires only that the approximate $g(r)$
vanish inside the cavity region. Because of the equivalence between the
structural predictions of the GFM and the HLR equation, these same
conclusions hold for the HLR equation. In particular, the density response
outside a tiny cavity is {\em exactly} described by linear response theory
about the uniform bulk system.

\subsection{Solvation free energy from thermodynamic pathways}

Moreover, these theories give the {\em exact} result of Eq.\ (\ref
{exact mu}) for the solvation free energy $\beta \Delta \Omega _{v}$,
independent of possible errors in $g(r)$, for all structurally based
thermodynamic pathways that use only tiny hard core fields. In particular,
the virial route in Eq.\ (\ref{mu virial final}) gives exact results for
$\beta \Delta \Omega _{v}$ since this requires only the (exact) contact
values $\rho _{\lambda }(\lambda R_{v}).$ The compressibility route as
written in Eq.\ (\ref{delomegachat}) requires $C({\bf {r}}_{1})$, which from
Eq.\ (\ref{gft c21}) depends only on the exact $\chi _{in}^{-1}$ in Eq.\ (\ref
{chiin small}). Both results require only that the bulk $g(r)$ vanish for
$r<\sigma ,$ and are unaffected by any errors at larger $r$.
This is in accord with our general supposition that
particular thermodynamic pathways can be relatively insensitive to errors in
the structural theory.

However the choice of pathway is important. Thus the density routes do not
give $\beta \Delta \Omega _{v}$ exactly even in the tiny cavity regime. This
is because as $\rho _{\lambda }({\bf {r}})$ is varied, the corresponding $
\phi _{\lambda }({\bf {r}})$ in general is not a pure hard core field and
spreads outside the cavity region. Neither the GFM nor the HLR theories can
treat these softer and longer-ranged fields exactly even if exact uniform
fluid correlation functions are used.

\subsection{Solvation free energy from the GFM partition function}

The $\beta \Delta \Omega _{v}$ obtained directly by taking the logarithm of
the partition function in Eq.\ (\ref{gft mu}) can be expressed analytically
for tiny cavities as
\begin{equation}
\beta \Delta \Omega _{v}=\frac{1}{2}\log {(1-\rho ^{B}}v{)}+\frac{1}{2}\frac{%
\rho ^{B}v}{1-\rho ^{B}v}.  \label{gft mu small}
\end{equation}
As $v\to 0$, this goes as $({\rho ^{B}}v)^{2}/4$, while the exact
result from Eq.\ (\ref{exact mu}) goes as ${\rho ^{B}}v$. This quadratic term
arises from the assumption of Gaussian fluctuations, which breaks down in
this limit. The alternate discrete version from Eq.\ (\ref{gft mu dis})
reduces to
\begin{equation}
\beta \Delta \Omega _v =-\log \frac{e^{-\bar{N}\,/\,[2(1-\bar{N})]}}{e^{-\bar{N}%
\,/\,[2(1-\bar{N})]}+e^{-(1-\bar{N})/[2\bar{N}]}},  \label{gft mu dis small}
\end{equation}
where $\bar{N}\equiv \rho ^{B}v.$ This has the peculiar behavior as $%
v\to 0$ that all derivatives vanish, and so is significantly in
error in this regime. See Fig.\ \ref{fig:rvsmall} for comparison with the
exact answer.

\section{Conclusion}

We have discussed several different thermodynamic routes that can be used
to determine the solvation free energy for inserting both small and large
cavities into a hard sphere fluid. Generally accurate results are found by
using the HLR equation to relate the densities and associated fields over
the intermediate states of the different pathways. We also considered the
GFM and showed that it gives results equivalent to the HLR equation for the
density response induced by a rigid cavity. However the GFM cannot describe
the softer external potentials and intermediate densities needed for the
density routes and for more general thermodynamic pathways. Direct use of
the approximate partition function of the GFM to determine the solvation
the free energy of a cavity also gives less accurate results.

Best results using the HLR equation for the solvation free energy of a
cavity are found from the compressibility and density routes. This can be
understood since most states along these routes require the density response
at intermediate densities and distances away from the cavity where the HLR
equation is most accurate. The HLR equation can also be used for
more general solutes with different shapes or longer ranged attractive
interactions and in applications where other pathways may be more
useful. Combined with an appropriate pathway it represents a versatile
and computationally efficient method for determining both the structure and
the thermodynamics of nonuniform fluids.

We thank Kirill Katsov, Michael Fisher, Jim Henderson, and Lawrence Pratt for
helpful comments and Gavin Crooks
for sending us the simulation data that is used in the plots of
finite size cavities. This work was supported by the National Science
Foundation through Grant CHE-0111104.

\appendix*
\section{}\nonumber
\label{surface tensions}
In the ``wall'' limit where $R_{v}\to \infty$, the surface tension given
by the compressibility route is determined from
\begin{equation}
\gamma _{\infty }=-\int_{0}^{\rho ^{B}}d\rho \frac{\partial \mu }{\partial
\rho }\int_{0}^{\infty }dz[\rho _w(z)-\rho ],  \label{gamma comp formal}
\end{equation}
with $\rho _w(z)\equiv \rho (z+R_{v})$. The virial route gives
\begin{equation}
\gamma _{\infty }=k_{B}T\rho ^{B}\int_{0}^{\infty }dR[G(R)-G(\infty )],
\label{gamma virial formal}
\end{equation}
Here $\rho ^{B}G(R)$ is the contact value of the density response to a
cavity of size $R$. In the case considered in this paper, the 
$\gamma_\infty$ from both
the compressibility and the virial routes can be determined analytically.
In general, when no analytical expressions are available, one may need
to carry out these integrations numerically.

There also exists the exact virial sum rule $\rho ^{B}G(\infty )=\beta p^B$
for a planar wall immersed in the hard sphere fluid \cite{henderson:statistical}.
However, this focuses on the structure at contact for the planar wall,
which is the region where the HLR equation is least accurate. Thus this
thermodynamic pathway for the bulk pressure gives relatively poor results.

\end{document}